# A SUCROSE SOLUTIONS APPLICATIION TO THE STUDY OF MODEL BIOLOGICAL MEMBRANES

M.A. Kiselev[*], P. Lesieur[#], A.M. Kisselev[*], D. Lombardo[#], M. Killany[*], S. Lesieur[%], M. Ollivon[%]

[*] - *Frank Laboratory of Neutron Physics, JINR, 141980 Dubna, Moscow reg., Russia*
[#] - *LURE, Bat. 209-D, B.P. 34, F-91898 Orsay, France*
[%] - *Pharmaceutical Faculty, University Paris-Sud, Chatenay Malabry F – 92296, France*

**ABSTRACT**

The small-angle X-ray and neutron scattering, time resolved X-ray small-angle and wide-angle diffraction coupled with differential scanning calorimetry have been applied to the investigation of unilamellar and multilamellar dimyristoylphosphatidylcholine (DMPC) vesicles in sucrose buffers with sucrose concentrations from 0 to 60%. Sucrose buffer decreased vesicle size and polydispersity and increased an X-ray contrast between phospholipid membrane and bulk solvent sufficiently. No influence of sucrose on the membrane thickness or mutual packing of hydrocarbon chains has been detected. The region of sucrose concentrations 30%-40% created the best experimental conditions for X-ray small-angle experiments with phospholipid vesicles.

PACS codes: 87.15.By, 87.15.Kg, 87.22.Bt, 61.10.Eq

Keywords: sucrose, sugars, phospholipid, contrast, X-ray scattering.

*Correspondence to* :
M.A. Kiselev, Frank Laboratory of Neutron Physics, Joint Institute for Nuclear Research, 141980 Dubna, Moscow region, Russia.
E-mail: kiselev@nf.jinr.ru;  fax: 7-096-21-65882; phone: 7-096-21-66977.

1. INTRODUCTION

The structure and properties of phospholipids membranes, as they serve as a simple and appropriate model for biological membranes, are investigated intensively by X-ray and neutron scattering techniques [1-4,8,9,16,19]. The structural investigation of drug delivery systems based on the lipid vesicles (liposomes) is a perspective direction for SR application [5]. Investigation of model biological membranes and mixed lipid/surfactant systems is one of the problems of interest [3,9]. The general problem of finding the universal solvent of suitable contrast for SR on highly diluted systems of biological importance is not solved yet. Great polydispersity in the sizes of the membrane vesicles in pure water constitutes another unsolved problem. In our previous synchrotron investigations of mixed phospholipid/surfactant systems, 40% sucrose/water solution was used as a contrast buffer [3]. New model for the evaluation of data from scattering on large multilamellar vesicles was developed recently [8]. This model describes the internal membrane structure by means of two strip functions, one corresponding to the region of polar head group and the other to the region of hydrocarbon chains. The introduction of a technique to create stable unilamellar vesicles with low size polydispersity can give a chance to upgrade the model in order to describe the internal membrane structure more precisely.

One of the important properties of sugars is possibility to stabilize the membranes and proteins during the drying. It is currently discussed in literature that sugars can replace the water molecules in dry lipid and protein systems [10]. Sugars (trehalose and sucrose) also decrease the vesicles leakage [11]. Dry vesicles can be stored in the sucrose glasses after the water evaporation [12].

The sucrose buffer has been applied for the investigation of micelle and vesicle structures by X-ray and neutron scattering [9]. The purpose of present work is to introduce the main results of the sucrose buffer properties investigation in the range of sucrose concentrations 0–60%.

2. EXPERIMENT

**D**imyristoylphosphatidylcholine (DMPC) - $C_{36}H_{72}NO_8P$, 8 Lauryl Ether ($C_{12}E_8$) - $C_{25}H_{58}O_9$ and sucrose - $C_{12}H_{22}O_{11}$ were purchased from Sigma (France). The unilamellar vesicles were prepared from 1% (w/w) of DMPC suspension in the sucrose/water solution by extrusion through polycarbonate filter with 500Å pore diameter as described in [1]. The multilamellar liposomes for the diffraction experiment were prepared in such a way that the number of solvent molecules per DPPC molecule was kept constant. Let us call $N_{SUC}$, $N_W$, $N_{lip}$ the number of sucrose, water, and DMPC molecules, respectively. For DMPC in pure water, the ratio $N_W / N_{lip}$ = 37.7 corresponds to 1:1 weight ratio. For DMPC in sucrose /water solvent the ratio $(N_{SUC} + N_W) / N_{lip}$ was kept equal to 37.7.

The small-angle X-ray scattering (SAXS) and X-ray diffraction measurements were carried out at D22 spectrometer of DCI synchrotron ring at LURE, France. The small-angle neutron scattering experiments (SANS) with unilamellar vesicles were carried out at YuMO spectrometer in Dubna, Russia [1,2,16]. The model of infinitely thin sphere was applied for the interpretation of the SAXS curves in the region of scattering vector q from 0.005Å$^{-1}$ to 0.04Å$^{-1}$ [3]. The macroscopic scattering cross section is given in this model by

$$\frac{d\Sigma}{d\Omega}(q) = n \cdot \left(4\pi \cdot R^2 \cdot d_l \cdot \Delta\rho\right)^2 \cdot \left(\frac{Sin(qR)}{qR}\right)^2, \qquad (1)$$

where **n** is the number of vesicles per unit volume, **R**-vesicle radius, **d$_l$**- membrane thickness, $\Delta\rho$ - X-ray contrast between membrane and aqueous sucrose solution. Due to the vesicles polydispersity, the macroscopic cross section was smeared with the introduced Gaussian distribution of vesicles radii, which gave the final expression for the vesicles macroscopic cross section **I$_c$(q)**

$$I_c(q) = \frac{\int_{R_o-3\sigma}^{R_o+3\sigma} \frac{d\Sigma}{d\Omega}(q,R) \cdot \exp\left(-\frac{(R-\bar{R})^2}{2\cdot\sigma^2}\right) \cdot dR}{\int_{R_o-3\sigma}^{R_o+3\sigma} \exp\left(-\frac{(R-\bar{R})^2}{2\cdot\sigma^2}\right) \cdot dR}, \qquad (2)$$

where σ is the standard deviation of **R**. The Gaussian distribution of vesicles radii was used as have been described in SANS study of egg yolk phosphatidylcholine vesicles [1].

The time resolved diffraction coupled with DSC was used to collect information from small-angle and wide-angle diffraction on the multilamellar membrane structures as described in detail in [4, 14]. The multilayered structure repeat distance **d** was determined from the position of the first diffraction peak using Bragg equation: $2 \cdot d \cdot \sin\theta = \lambda$, where $\theta$ is a half of the scattering angle.

## 3. RESULTS AND DISCUSSION

The X-ray scattering density of DMPC molecule in the $L_\alpha$-phase, $\rho_{DMPC} = 0.957 \cdot 10^{11} cm^{-2}$. The X-ray scattering density of sucrose buffer is calculated by the expression

$$\rho_{solvent} = 16.979 \cdot \left(\frac{1-\chi}{\chi} + 0.0532 \cdot \chi\right) \cdot D(\chi) \cdot 10^{11}, cm^{-2} \qquad (3)$$

where, χ - sucrose concentration (w/w), **D(χ)** - density of sucrose/water solution in g/cm$^3$ [15]. $\Delta\rho = \rho_{DMPC} - \rho_{solvent}$ is a contrast between membrane and solvent. The results of contrast calculations are shown in Fig. 1.

In SAXS experiments the sucrose concentration range from 0 to 50% (w/w) was chosen because of the limited viscosity increase in this domain. The SAXS spectra from each sample were collected during 20min. The experimental macroscopic cross sections of DMPC vesicles at T=30°C with sucrose concentrations 0%, 20%, 30%, 45% are presented in Fig. 2. The spectra for the sucrose concentrations 20%, 30%, 45% were fitted well by the model of infinitely thin sphere (Eq. 1) with Gaussian distribution of vesicle sizes (Eq. 2). The scattering curve from vesicles in pure water has too poor statistics to be interpreted. The macroscopic cross section of vesicles increases with the increase of sucrose concentration due to the increase in the contrast as shown in Fig. 1. At sucrose concentration 15%, the statistical errors decreased down to the values that give opportunity for the model application. The results for sucrose concentration in the interval 15-50% are presented in Table 1. The average radius of vesicle and polydispersity have their minimum values in the range of 30-40% sucrose. This interval exhibits the best experimental conditions for the study of vesicles structure by X-ray scattering.

Fig. 1 presents the comparison between theoretical values of the contrast and contrast values determined experimentally from the measured values of macroscopic cross-section, $\bar{R}$, σ, and **$d_l$**, according to the Eqs. (1) and (2). The membrane thickness was determined from SANS experiment, which gives a constant value of the membrane thickness **$d_l$**=38.8±0.8Å at T=30°C for the sucrose concentration in the range 0-40% [1,2,16]. This value was used in the calculations of the contrast from the experimentally measured SAXS curves. The values determined by different methods are in good agreement.

Other solvents, such as glycerol and dimethyl sulfoxide (DMSO), can increase the contrast for the phospholipid membranes relative to that for water. At T=30°C, the electron density of $H_2O$ is $3.35 \cdot 10^{23} e/cm^3$, the electron density of glycerol - $3.60 \cdot 10^{23} e/cm^3$, the electron density of DMSO - $3.56 \cdot 10^{23}$e/cm$^3$, the electron density of 20% sucrose solution is $3.58 \cdot 10^{23}$e/cm$^3$. But the glycerol influences the membrane structure and at 0.5 molar fraction creates the interdigitated phase [17]. Our SANS experiments with phospholipid unilamellar vesicles in the DMSO/water solutions have demonstrated that starting from 0.1-0.15 molar DMSO fraction vesicles began to aggregate and create the multilamellar structures. So, neither glycerol nor DMSO can be used as a buffer media to study the vesicles structure by SAXS.

The second important question to be clarified is the sucrose influence on the multilamellar structure of DMPC. The DSC curves recorded are presented in Fig. 3. The first endothermic peak corresponds to the transition from gel $L_{\beta'}$-phase to ripple $P_{\beta'}$-phase, so-called pretransition. For 0% sucrose concentration the peak maximum is located at about 15°C. The second endothermic peak corresponds to the transition from ripple $P_{\beta'}$-phase to liquid crystalline $L_\alpha$-phase, so-called main phase transition. The onset temperature of main phase transition $T_m$=22.8°C for DMPC in pure water. The DSC results are presented in Table 2. The pretransition temperature is shifted towards the main phase transition temperature with the increase of sucrose concentration. The enthalpy of the main phase transition does not change with the increase of sucrose concentration up to 40%. At 60% concentration the enthalpy decreases to the value of 6.0cal/g. It was measured by C. Fabrie that the enthalpy of DMPC main phase transition decreases linearly from the value of 8.1cal/g at 0% sucrose to the value of 7.3cal/g at 60% sucrose [11]. The value of the main phase transition temperature from our measurements is in good agreement with C. Fabrie's results. In the interval of experimental errors, our results for the enthalpy of the main phase transition are in the agreement with results of C. Fabrie for sucrose concentration from 0% to 40%, and are different at sucrose concentration 60%. The discrepancy in the values of enthalpy cause from the overlap of the signals from the pretransition and main phase transition at 60% sucrose.

The SAXS and WAXS diffraction patterns were recorded with acquisition time of 1 min at the temperature alteration from 3°C to 38°C. It allows one to calculate the repeat distance of DMPC membrane as a function of temperature. SAXS and WAXS patterns from DMPC multilamellar liposomes are presented in Figs. 4 and 5 for 0%, 20%, 40%, 60% sucrose at T=10°C. The systems studied were in gel $L_{\beta'}$-phase at this temperature. The thickness of the intermembrane solvent space was calculated as $d_s=d-d_l$. The value of $d_l$ was determined from SANS experiment with large unilamellar vesicles as described in [1,2,16]. Sucrose has no influence on $d_l$ so it has a permanent value of 44.5±1.0Å at T=10°C. The addition of 20% sucrose increases $d_s$ from 15.1±1.9Å to the value of 20.4±2.0Å, which does not change much for higher sucrose concentrations, see Table 3.

The WAXS diffraction from the hydrocarbon chains of DMPC molecules in $L_{\beta'}$-phase (Fig. 5) demonstrates that the packing of hydrocarbon chains has a permanent structure for all sucrose concentrations. The WAXS patterns exhibit sharp (2,0) reflection at q=1.496±0.003Å$^{-1}$ and broad (1,2) shoulder at q=1.52±0.03Å$^{-1}$ for all sucrose concentrations. These reflections correspond to the disordered hexagonal lattice, which can be represented as two-dimensional rectangular lattice (containing 4 hydrocarbon chain projections ) of dimensions a=8.40±0.17Å, b=9.49±0.28Å [18,4]. This lattice corresponds to an area per hydrocarbon chain $S_{ch}$=19.9±1.0Å$^2$. WAXS diffraction patterns in the ripple phase have broad diffraction peak at q=1.50±0.03Å$^{-1}$ (rectangular lattice with a=8.38±0.17Å, b=9.67±0.19Å) for all sucrose concentrations. The area per hydrocarbon chain $S_{ch}$=20.4±0.8Å$^2$. In the liquid crystalline $L_\alpha$-phase the diffraction peak becomes broader and has a permanent maximum at $q=1.43\pm0.03$Å$^{-1}$ for all sucrose concentrations (rectangular lattice with $a=8.80\pm0.18$Å, $b=10.16\pm0.21$Å, $S_{ch}=22.4\pm0.9$Å$^2$). The increase in the area occupied by hydrocarbon chain projection on the membrane surface is a result of an increase in the hydration of polar head groups in the $L_\alpha$-phase. Consequently, no influence of sucrose on the hydration of polar head groups was detected.

The main result from WAXS diffraction is that the packing of hydrocarbon chains is not influenced by the sucrose in all studied phases, which supports the conclusion about permanent membrane thickness in sucrose buffers from SAXS and SANS experiments. The present WAXS results are in agreement with WAXS results published by J. Stumpel in [13] for the liquid crystalline $L_\alpha$-phase, but are different for the case of $L_{\beta'}$-phase. J. Stumpel reported that the spacing between hydrocarbon chains is a function of sucrose concentration with minimal lattice at 30% sucrose. Above 50% sucrose the WAXS diffraction peak becomes symmetrical, which was explained in terms of change in the tilt of hydrocarbon chains. Our results demonstrate that at 60% sucrose (see Fig. 5) peak has the same shoulder as at 0% sucrose concentration.

The diffraction pattern for the ripple $P_{\beta'}$-phase consists of two types of peaks (see Fig. 6). The first type is the diffraction from multilamelar membrane structure, the first order sharp diffraction peak at about $q=0.1$Å$^{-1}$ for 0% sucrose. The second type is the diffraction on the undulations of the membrane surface, the first order peak for 0% sucrose located at $q=0.052\pm0.002$ Å$^{-1}$. The value of undulation length $d_r$ is calculated as $d_r =2\pi/q$. The values of $d_r$ are 120.8±3.6Å for sucrose concentration 0%, 120.8±3.6Å for 20% sucrose, and 128.5±7.7Å. for 40% sucrose, i.e. the membrane undulation parameter has the same value for all sucrose concentration from 0% to 40%.

The diffraction peaks of first and second order from multilamellar membrane structure in the ripple phase consist of two slightly overlapping diffraction peaks. In Fig. 6 the peak splitting can be seen only in the second diffraction peak from multilamellar membrane structure for 0% sucrose concentration. The peak splitting at 0% sucrose corresponds to the values of $d$ 64.6±1.0Å and 63.5±1.0Å. At the 20% sucrose two well distinguished diffraction peaks from multilamellar membrane structure are detected, corresponding to the values of $d$ 74.8±1.1Å and 68.3±1.0Å. At 40% sucrose concentration, the broad diffraction peak corresponds to $d=70.5\pm1.1$Å. The diffraction peak position cannot be determined at sucrose concentration 60%. The diffraction intensity decreases due to the decrease in the number of layers in the multilamellar liposomes. This behavior is typical for all phases studied (see Figs. 4,6) and can explain the fact established by the SANS that vesicles in sucrose buffer are sufficiently stable in time compared to the pure water buffer. It is possible to say that sucrose posses weak surfactant properties.

In the ripple phase, the value of $d$ increases with the increase of sucrose concentration, as it is seen from comparing $d$ values for sucrose concentrations 0%, 20%, 40%. No influence of sucrose on the membrane undulations and, consequently, on the membrane rigidity was found.

In the liquid crystalline $L_\alpha$-phase (figures not shown), the diffraction peak from multilamellar membrane structure with $d=62.8\pm0.9$Å at 0% sucrose and T=30°C is split up into two diffraction peaks. At 20% sucrose they correspond to the values of repeat distances 61.0±0.9Å and 68.3±1.0Å. At 40% sucrose, the difference in the value of repeat distances becomes smaller and harder to distinguish, the values of $d$ are 62.2±0.9Å and 64.2±1.0Å at T=31°C. The peak splitting at 40% sucrose concentration was measured in $L_\alpha$-phase for diluted system with concentration of DMPC 5% (w/w). Despite the very low statistics, the splitting of diffraction peak was detected. The membrane exhibits two phase coexistence with lamellar spacing 58.7±2.0Å and 75.7±2.0Å. The two phase coexistence was not detected by the measurements of J. Stumpel at T=35°C with very diluted system. The concentration of DMPC was 0.5% (w/w) [13]. These results show one phase with increasing repeat distance from 62±0.5Å at 0% sucrose to 66.5±0.5Å at 14% sucrose and with further monotonous decrease to the value of 58.5±0.5Å at 60% sucrose.

The liquid $L_\alpha$-phase of DMPC has a complex structure in the presence of sucrose, which depends on the sucrose and DMPC concentrations in the buffer. The properties of multilamellar DMPC membranes in sucrose buffers cannot be explained as a simple penetration of sucrose molecules into the region of polar head groups. There are strong suggestions in the literature that sucrose, either free in solution or covalently linked to membrane surface, can also affect the physical properties of the membrane. Probably the sucrose influence on the membrane is not direct, at first sucrose influences the buffer properties and through that the membrane [10]. Our results support this idea. Sucrose has no influence on the membrane thickness of unilamellar vesicles, but has an influence on the polydispersity and vesicle radius. Sucrose molecules do not influence the mutual packing of hydrocarbon chains, but increase the main phase transition temperature and influence the intermembrane interaction by increasing the repulsive forces.

## 4. CONCLUSIONS

As the result of complementary synchrotron and neutron scattering experiments, it is established that aqueous sucrose solutions: have no influence on the membrane thickness and the mutual packing of hydrocarbon chains; have an influence on the properties of phase transitions. The membrane thickness has a constant value in the range of sucrose concentrations 0%-40% (44.5±1.0Å at T=10°C). Sucrose buffer sufficiently increases X-ray contrast for model biological membranes, decreases the vesicles polydispersity, and increases the vesicles life-time. It gives a possibility to determine the average value of vesicles radius.

Sucrose buffer in the range of concentrations 30%–40% is a perspective medium for the SAXS application to the investigation of the structure of vesicles and mixed lipid/surfactant aggregates. In this respect, the structural synchrotron investigations of phospholipid based drug carriers in the range of vesicle sizes 500Å-1000Å have a special practical interest.

**ACKNOWLEDGEMENTS**
The experiments at LURE was supported by the TMR program for Great Instruments. The authors are grateful to Dr. L.I. Barsukov (Institute of Bioorganic Chemistry, Moscow) for the help with sucrose buffer preparation and to Prof. P. Balgavy (Faculty of Pharmacy, Comenius University, Bratislava) for the fruitful discussions.

Table 1.
The parameters of spherical model with infinitely thin surface. **R** - average radius of sphere, $\sigma$-standard deviation of **R**, **P=1.18·$\sigma$/R** - polydispersity. $C_{suc}$ - sucrose concentration (w/w) in the buffer.

| $C_{suc}$, % | 15    | 20    | 25    | 30    | 35    | 40    | 45    | 50     |
|--------------|-------|-------|-------|-------|-------|-------|-------|--------|
| **R**, Å     | 228±2 | 237±1 | 213±1 | 218±1 | 216±1 | 218±1 | 216±1 | 130±1  |
| **$\sigma$**, Å | 83±2  | 74±1  | 78±1  | 61±1  | 60±1  | 61±1  | 77±1  | 105±10 |
| **P**, %     | 36    | 31    | 33    | 28    | 28    | 28    | 36    | 81     |

Table 2.

The dependence of DMPC main phase transition temperature $T_m$, pretransition temperature $T_p$, and enthalpy of main phase transition $\Delta H_m$ on the sucrose concentration $C_{suc}$.

| $C_{suc}$, % | 0 | 20 | 40 | 60 |
|---|---|---|---|---|
| $T_m$, Å | 22.8±0.2 | 23.2±0.2 | 24.2±0.2 | 25.1±0.2 |
| $T_p$, Å | 14.6±0.2 | 15.7±0.2 | 18.3±0.2 | 22.8±0.2 |
| $\Delta H_m$, cal/g | 7.9±0.5 | 8.3±0.5 | 7.9±0.5 | 6.0±0.4 |

Table 3.

The dependence of DMPC membrane repeat distance $d$ and thickness of the intermembrane solvent $d_S$ on the sucrose concentration $C_{suc}$ in $L_{\beta'}$-phase (T=10°C ).

| $C_{suc}$, % | 0% | 20% | 40% | 60% |
|---|---|---|---|---|
| $d$, Å | 59.6±0.9 | 64.9±1.0 | 65.1±1.0 | 65.5±3.0 |
| $d_S$, Å | 15.1±1.9 | 20.4±2.0 | 20.6±2.0 | 21.0±4.0 |

**FIGURE LEGENDS**

Fig. 1. The contrast $\Delta\rho$ between DMPC membrane in the $L_\alpha$ phase and bulk solvent. Squares, theoretically calculated values. Circles, the values of contrast determined from SAXS experiment.

Fig. 2. SAXS curves for extruded DMPC vesicles in the aqueous solutions with sucrose concentrations 0% (circles), 20% (squares), 30% (triangles), 45% (rhombuses) at T=30°C. The fitted curves are presented (solid curves).

Fig. 3. DSC curves for 0%, 20%, 40%, 60% sucrose concentrations recorded at heating rate 1°C/min. At 0% sucrose DSC signal in voltage, all other curves are shifted along ordinate axis. The first endothermic peak corresponds to the pretransition and the second - to the main phase transition

Fig. 4. SAXS diffraction patterns from multilamellar liposomes at T=10°C ($L_{\beta'}$-phase) with 0%, 20%, 40%, 60% sucrose concentrations. The diffraction patterns are shifted along ordinate axis.

Fig. 5. WAXS diffraction patterns from spatial packing of hydrocarbon chains of lipid molecules at T=10°C ($L_{\beta'}$-phase). The diffraction peaks have permanent positions, sharp peak at $q$=1.50Å$^{-1}$ with broad shoulder at $q$=1.52Å$^{-1}$. The diffraction patterns are shifted along ordinate axis.

Fig. 6. SAXS diffraction patterns from multilamellar liposomes in the ripple phase with 0%, 20%, 40%, 60% sucrose concentrations. **T** in the range of 21°C - 23°C. For 0% sucrose, the diffraction peaks at $q$=0.052 Å$^{-1}$ and shoulder at $q$=0.108Å$^{-1}$ are first and second order diffraction from the membrane undulations. The diffraction patterns are shifted along ordinate axis.

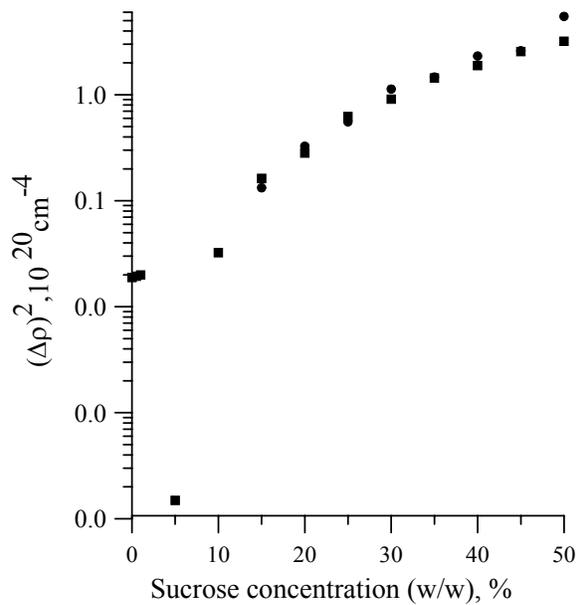

Fig. 1

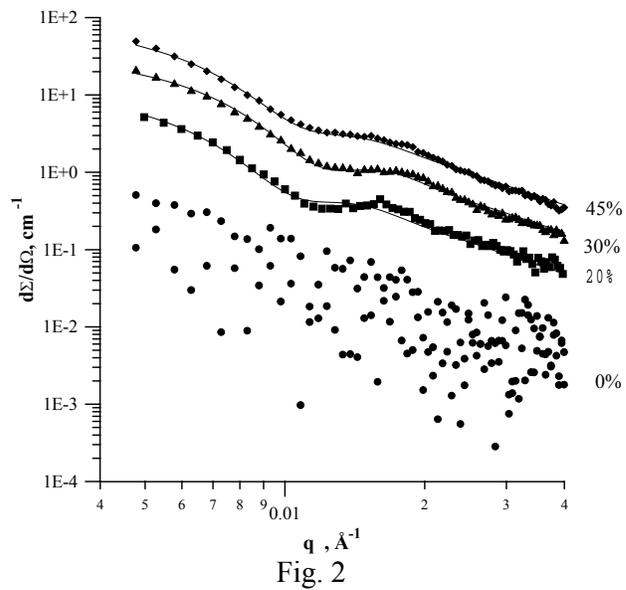

Fig. 2

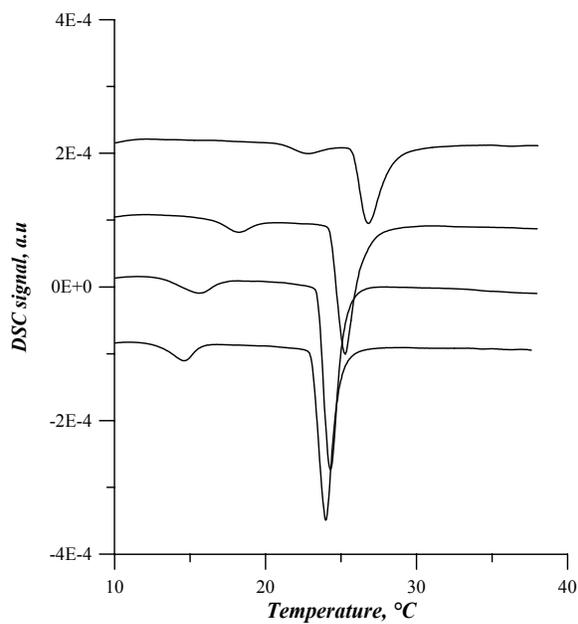

Fig. 3

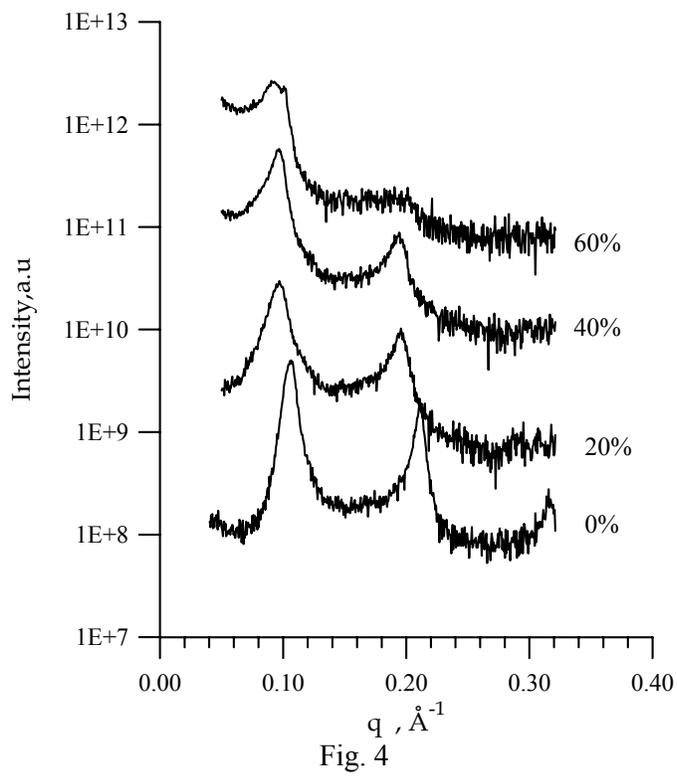

Fig. 4

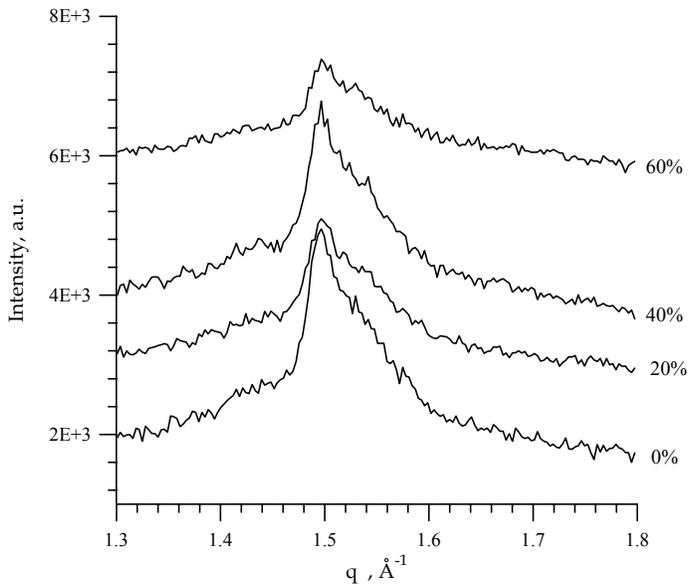 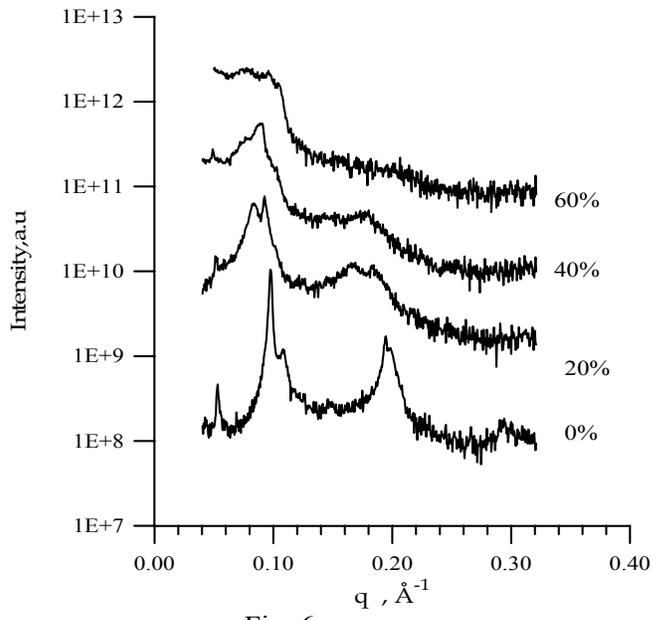

Fig. 5.  Fig. 6.